\documentclass{aa}
\begin{document}

   \thesaurus{05     
              (05.01.1;  
               08.09.2 $\upsilon$ And; 
               08.16.2; 
               03.19.2; 
               03.13.2)}  

   \title{Detection and measurement of planetary systems with GAIA}

   \author{A. Sozzetti\inst{1,3}
          \and
          S. Casertano\inst{2}
          \and
          M. G. Lattanzi\inst{3}
          \and
          A. Spagna\inst{3}
          }

   \offprints{A. Sozzetti (alex@phyast.pitt.edu)}

   \institute{The University of Pittsburgh, Department of Physics and
   Astronomy, Pittsburgh, PA 15260, USA
         \and
              Space Telescope Science Institute, Baltimore, MD
              21218, USA
         \and
              Osservatorio Astronomico di Torino, 10025 Pino Torinese, Italy
             }

   \date{Received ???; accepted ???}
   \authorrunning{A. Sozzetti et al.}
   \titlerunning{Measurement of planetary systems with GAIA}
   \maketitle

   \begin{abstract}

    We use detailed numerical simulations and the $\upsilon$
Andromed\ae \, planetary system as a template to evaluate the
capability of the ESA Cornerstone Mission GAIA in detecting and
measuring multiple planets around solar-type stars in the
neighborhood of the Solar System. For the outer two planets of the
$\upsilon$ Andromed\ae \, system, GAIA high-precision global
astrometric measurements would provide estimates of the full set
of orbital elements and masses accurate to better than 1--10\%,
and would be capable of addressing the coplanarity issue by
determining the true geometry of the system with uncertainties of
order of a few degrees. Finally, we discuss the generalization to
a variety of configurations of potential planetary systems in the
solar neighborhood for which GAIA could provide accurate
measurements of unique value for the science of extra-solar
planets.

   \end{abstract}

%

\section{Introduction}

As of today, five normal stars in the solar neighborhood are known
to harbor either two or more planets
(\cite{butler,mayor,marcy,udry}) or systems composed of a planet
and a brown dwarf (\cite{marcy2,udry,els}). The observational
evidence of the first extra-solar planetary systems has
immediately raised crucial questions regarding their formation and
evolution. Are the orbits coplanar? Are the configurations
dynamically stable? The quasi-coplanarity hypothesis is one of the
dogmas in present planet formation theories~(\cite{lissa93}). For
example, in the case of the three-planet system around the F8V
star $\upsilon$ Andromed\ae\, ($\upsilon$ And hereafter), Laughlin
\& Adams~(\cite{laughlin}) and Stepinski et al.~(\cite{stepinski})
have shown that orbital evolution and long-term stability of the
system can be greatly affected by small variations in the
three-companion orbital model. In particular, studies on the
dynamical stability of the outer two companions require
constraints to be placed on the maximum allowed values for the
masses and on the range of allowed relative inclinations of the
two orbits. Radial velocity measurements cannot determine either
the inclination $i$ of the orbital plane with respect to the plane
of the sky or the position angle $\Omega$ of the line of nodes in
the plane of the sky: without knowledge of the full set of orbital
parameters and true mass values, general conclusions on the
architecture, orbital evolution and long-term stability of the
newly discovered planetary systems may be questionable.

In this Letter, we gauge the potential of GAIA in detecting and 
measuring multiple planets in favorable cases, i.e. well-spaced,
well-sampled orbits, and high ``astrometric'' $S/N$ ratios (see
Sect.~3). In order to do so, we focus on the $\upsilon$ And
system, which, given its characteristics, constitutes a
well-suited laboratory. In this exploratory Letter, we do not 
address issues such as insufficient orbital
sampling, resonant configurations, and astrometric signatures
close to the detection limit, which clearly need to be considered 
in a more realistic simulation. Work on these issues is in 
progress and will be presented in the future.

\section{Data simulation and analysis method}

The code for the reproduction of GAIA global astrometric
observations and analysis of the simulated dataset has been
thoroughly described in our previous work~(\cite{lattanzi00}),
where we defined detectability horizons and limits on distance for
accurate orbital parameters and mass determination in the case of
single Jupiter-mass planets orbiting single solar-type stars.

We have modified our simulation code to allow for the presence of
multiple planets around each target star (\cite{casertano00}).
The stellar motion is described in terms of the 5 basic
astrometric parameters and of the gravitational perturbations
produced by the orbiting planet(s), which are assumed to be linear
(i.e., as the sum of independent Keplerian motions). Such
simplification (no mutual interaction terms between planets) has
little impact on the significance of the analysis. In fact, on the
time-scale of GAIA observations (5 years), variations of the
orbital parameters can be confidently considered negligible. For
planet detection we employ a standard $\chi^2$-Test with
confidence level set to 5\% applied to the observation
residuals~(\cite{lattanzi00}), while in the case of multiple
orbits the iterative method for the solution of the non-linear
systems of equations of condition is based on a slightly modified
version of the Levenberg-Marquardt algorithm~(\cite{casertano99}),
which ensures stability of the solution.

\section{GAIA observations of the $\upsilon$ And system}

In addressing the problem of the detection and measurement of
multiple planets with GAIA, we have utilized the $\upsilon$ And
planetary system as a representative test case. The results are
discussed below.

\subsection{Detection}

GAIA (for a detailed review of the mission concept
see~\cite{gilmore}) surveys the sky utilizing a Hipparcos--like
scanning law. Its elementary uni-dimensional measurement is the
longitude $\psi$ of the Reference Great Circle being scanned at a
given time. The single-measurement astrometric error $\sigma_\psi$
was fixed to 10 $\mu$as throughout all our simulations: this value
applies to stars brighter than $V=12$ mag. Sensitivity to
planetary perturbations is driven by the astrometric error
$\sigma_\psi$ of each measurement, which has to be compared to the
magnitude of the gravitational perturbation induced on the parent
star by one or more orbiting planets: the {\it astrometric
signature} $\alpha$ (in arcsec) is the apparent amplitude of the
orbital motion of the star around the system barycenter, defined
as
\begin{equation}
  \alpha = \frac{M_\mathrm{p}}{M_\star}\frac{a_\mathrm{p}}{d},
\end{equation}
where $M_\mathrm{p}$, $M_\star$ are the masses of the planet and
star respectively (in solar masses), $a_\mathrm{p}$ the semi-major
axis of the planetary orbit (in AU), $d$ the distance of the
system from the observer (in pc).
\begin{table}
\caption{$\upsilon$ Andromed\ae: stellar parameters and orbital
elements utilized in the simulations. Planet masses and
astrometric signatures are computed as lower limits corresponding
to edge-on configurations ($\sin i = 1$)} \label{upsparam}
\begin{center}
\begin{tabular}{cccc}
\hline\hline \noalign{\smallskip} \multicolumn{2}{c}{Parameters} &
\multicolumn{2}{c}{$\upsilon$ And} \\\noalign{\smallskip} \hline
\multicolumn{2}{c}{$\lambda$ (deg)}& \multicolumn{2}{c}{37.9} \\
\multicolumn{2}{c}{$\beta$ (deg)}& \multicolumn{2}{c}{28.9} \\
\multicolumn{2}{c}{$\mu_\lambda$ (mas/yr)}&
\multicolumn{2}{c}{$-$242.74} \\ \multicolumn{2}{c}{$\mu_\beta$
(mas/yr)}& \multicolumn{2}{c}{$-$266.98} \\
\multicolumn{2}{c}{$\pi$ (mas)}& \multicolumn{2}{c}{74.25} \\
\multicolumn{2}{c}{Visual Magnitude}& \multicolumn{2}{c}{4.09} \\
\multicolumn{2}{c}{Spectral Type}& \multicolumn{2}{c}{F8V} \\
\multicolumn{2}{c}{$d$ (pc)}& \multicolumn{2}{c}{13.47} \\
\multicolumn{2}{c}{$M_\star$ ($M_\odot$)}& \multicolumn{2}{c}{1.3}
\\\noalign{\smallskip}\hline \noalign{\smallskip}
Orbital elements & $\upsilon$ And B & $\upsilon$ And C &
$\upsilon$ And D \\ \noalign{\smallskip}\hline
\noalign{\smallskip} $\alpha$ ($\mu$as)&2.40&100.1&658.4 \\$a$
(AU) &0.059&0.83&2.50
\\$P$ (d) &4.6&241.2&1266.6 \\ $e$ &0.034&0.18&0.41 \\ $\tau$
(JD)&24500088.6&2450154.9&2451308.7
\\$\omega$ (deg)&83.0&243.0&247.7\\$M_\mathrm{p}\sin i$
($M_\mathrm{J}$)&0.71&2.11&4.61 \\\noalign{\smallskip} \hline
\end{tabular}
\end{center}
\end{table}

We simulate 200 planetary systems on the sphere using the
parameters of the $\upsilon$ And system in the original paper by
Butler et al. (1999), as summarized in Table~\ref{upsparam}. Two
orbital elements are not determined by radial velocity
measurements, namely $i$ and $\Omega$. We simulate systems with 
random uniform distribution in $\Omega$ and express detection
probabilities as a function of the inclination of the orbital
plane ($5^\circ\leq i\leq 90^\circ$).\footnote{Astrometric detection
probability does not depend on the apparent orientation of the
orbit~(\cite{lattanzi00})} We assume the orbits of the three
planets are perfectly coplanar, i.e. $i$ and $\Omega$ are exactly
the same for all the system components. Planet masses are scaled
with the co-secant of the inclination angle. During the 5-year
mission lifetime, the system is observed for a total of 70
times\footnote{The GAIA scanning law is such that the number of
basic observations is function of the ecliptic latitude $\beta$}.

We find that, given the favorable values of the ``astrometric''
{\it signal-to-noise ratio} $S/N = \alpha/\sigma_\psi$, the
presence of both the outer and middle planet can be easily
detected (at the 95\% confidence level) by GAIA (in the residuals
to a single--star model and to a model containing the outer
companion, respectively), regardless of the inclination of the
orbital plane. In particular, the system configuration is such
(orbits with well sampled periods, high $S/N$, no resonances) that
detectability of the middle planet is not affected by the presence
of the outermost companion. On the other hand, the inner planet,
$\upsilon$ And B, cannot be detected, unless $i\leq 10^\circ$: as
a matter of fact, its orbital period, shorter than the average
observing period, is poorly sampled by GAIA, and most importantly
the magnitude of the astrometric signature is such that, for
$i\geq 10^\circ$, $S/N \leq 1$ and detection through the 
$\chi^2$-Test fails.

\subsection{Orbit reconstruction}

We focus here on GAIA's ability to make accurate measurements of
the orbital elements and masses for the outer two planets of the
system, which produce the greatest astrometric perturbations.
Specifically, we have simulated systems containing the middle or
the outer planet only (utilizing a single-planet solution for the
orbital fits), and the complete configuration with the three
planets, fitting to the observation residuals $a)$ a two-planet
model, disregarding the presence of the gravitational perturbation
induced by the inner planet, or $b)$ the complete three-planet
model (for details on the fitting procedure,
see~\cite{casertano99} or~\cite{sozzetti00}).

\begin{figure}[bh]
  \centering
  \vspace{8.5cm}
  \includegraphics{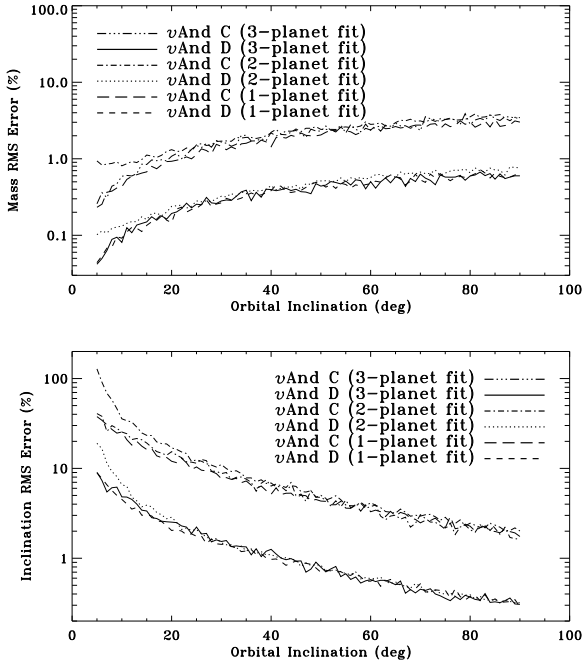}
  \caption{RMS errors for $M_\mathrm{p}$ and $i$, in the case of the outer
  two planets of the $\upsilon$ And system, expressed as fraction (\%) of the
  true values of the parameters: for $\upsilon$ And D, mass can be measured to
  0.1--1\% accuracy and inclination to better than 1\% (for $i\geq
  30^\circ$); for $\upsilon$ And C, mass is accurate to 1--3\% and inclination
  to better than 10\% (for $i\geq 30^\circ$)}
  \label{upsrms}
\end{figure}

In Figure~\ref{upsrms} we have summarized the results for two of
the most important parameters of the system, inclination $i$
(directly measured) and planet mass (computed once a reliable
estimate for the mass of the parent star is provided), for
$\upsilon$ And C and $\upsilon$ And D. Owing to the difference in
astrometric signature (see Table~\ref{upsparam}), the outermost
planet is measured more accurately than the inner planet (accuracy
is almost an order of magnitude higher).\footnote{Note that the
multiple-planet fit provides very accurate estimates for all of
the orbital elements of the outer two planets.} As the inclination 
$i$ decreases, towards a face-on configuration, the true mass 
inferred for each planet from the radial velocity data increases, 
therefore the astrometric signature is larger and provides a more 
accurate measurement of the planet's mass. At the same time, the 
observations are less sensitive to the inclination itself, and its 
fractional measurement error increases. It should also be noted that, 
if the system is very close to face-on, the accuracy of the 
inclination measurement will be substantially increased by combining 
radial velocity and astrometric data.

Finally, we emphasize the fact that, similarly to what happened
for detection, the orbital arrangement of the $\upsilon$ And
system is such that it does not affect our ability to
reconstruct each planet's orbit. As shown in
Figure~\ref{upsrms}, the two outer planets can have orbital
elements and masses measured almost as well as if each of them was
the only orbiting companion. For a two-planet orbital fit to a
complete three-planet simulation, the quality of orbit
reconstruction degrades only for $i\leq 10^\circ$, when the signal
from $\upsilon$ And B becomes greater than the single-measurement
error (we recall $\sigma_\psi = 10$ $\mu$as) , and thus detectable
in the observation residuals.

\subsection{Coplanarity analysis}

\begin{figure}[b]
  \centering
  \vspace{9.5cm}
  \includegraphics{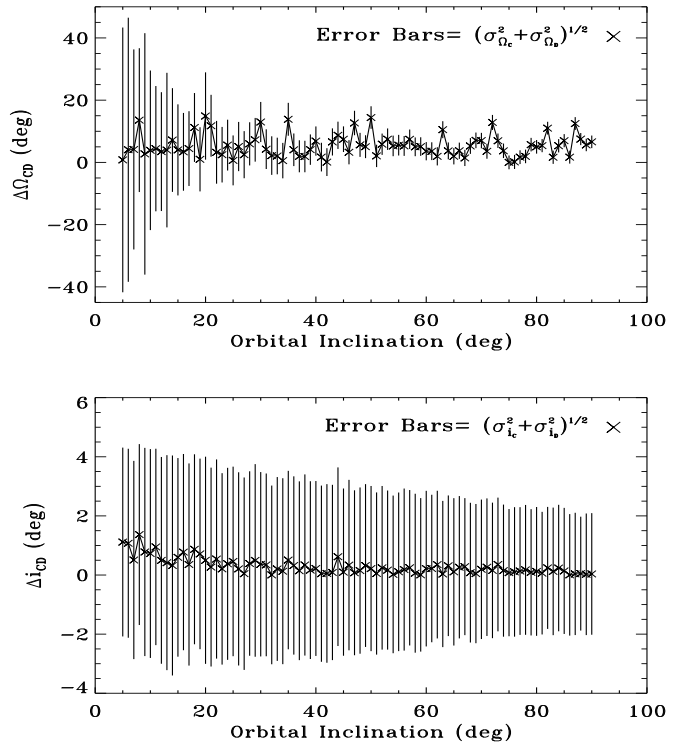}
  \caption{Coplanarity analysis for the $\upsilon$ And system:
  for $i > 20^\circ$, the orbits of the two outer planets can be confidently
  identified as quasi-coplanar, with (absolute) uncertainties of order of a
  few degrees (errors are computed using the formal expressions from the
  covariance matrix of the multiple-planet fit)}
  \label{upscoplan}
\end{figure}
As discussed in the Introduction, the unexpected orbital
configuration of the $\upsilon$ And system has motivated detailed
theoretical studies on its dynamical evolution and long-term
stability. For example, neglecting to first order the effects of
the innermost planet on the overall stability of the system,
Stepinski et al.~(\cite{stepinski}) come to the conclusion that
dynamical stability requires the orbital inclination of the outer
two companions to be greater than $i\sim 13^\circ$, otherwise the
two objects would be too massive and gravitational interactions
would disrupt the system. Furthermore, the system cannot be stable
in the long term if relative inclinations are greater than
$55^\circ$, $35^\circ$, and $10^\circ$, for $i\sim 64^\circ$,
$i\sim 30^\circ$, and $i\sim 15^\circ$, respectively: then, the
more massive the planets, the closer to coplanarity their orbits
have to be, for the system not to be destabilized on a short
time-scale.

Figure~\ref{upscoplan} shows the estimated accuracy with which
GAIA would measure the coplanarity of the orbits of $\upsilon$ And
C and $\upsilon$ And D (as shown in Sect.~3.1, unless its orbit is
almost face-on, $\upsilon$ And B is not even detectable). Fitting
a three-planet model to the simulated observations, we estimate
differences between position angles of the lines of nodes
($\Delta\Omega_{\mathrm{CD}}$) $\leq 10^\circ$ and differences
between inclinations of the orbital planes ($\Delta
i_{\mathrm{CD}}$) $\leq 3^\circ$, for $i > 20^\circ$. Thus,
quasi-coplanarity can be reliably established. As $i$ decreases,
the uncertainty on $\Delta\Omega_{\mathrm{CD}}$ grows, because of
the increasing difficulty in the correct identification of the
position angle $\Omega$, which becomes undefined at $i = 0^\circ$.
Simulations of non-coplanar systems with
$\Delta\Omega_{\mathrm{CD}}\neq 0^\circ$ and/or $\Delta
i_{\mathrm{CD}}\neq 0^\circ$, yielded similar results. Our
findings indicate that GAIA high-precision measurements would be
instrumental in verifying the system stability, except for
quasi-face on orbits: however, the inclination of $\upsilon$ And D
is probably greater than $10^\circ$--$15^\circ$, otherwise the
astrometry of Hipparcos would have revealed its
presence~(\cite{mazeh}).

\section{Discussion and conclusions}

The ability to accurately measure orbital parameters and masses
for a potentially large number of planetary systems with massive
planets is the key element of the GAIA contribution to the science
of extra-solar planets.

Utilizing the $\upsilon$ And three-planet system as a template, we
have shown how, in case of planetary systems with planets
producing favorable astrometric signatures ($S/N\geq 10$) and with
well-sampled orbital periods (shorter than the 5-year mission
lifetime), the parameters and masses of each companion can be
measured almost as well as if it was the only planet.

We have made coplanarity tests, and found that GAIA would be able
to determine whether orbits are coplanar or not with uncertainties
of a few degrees, thus providing theory with the observational
evidence needed to address the long-term stability issue for the
system .

Lattanzi et al. (2000) derive a 60--pc distance horizon for 30\%
accuracy measurements of orbital elements and masses in the case
of {\it single} Jupiter-mass planets orbiting solar--type stars
with periods ranging between 0.5 and 11.8 years (the true period
of Jupiter).

We can generalize those findings, in an early attempt to identify
the variety of possible configurations of planetary systems
detectable and measurable with GAIA. The results reported in this
letter allow us to say that the 60--pc limit on distance holds for
detection and measurement accurate to 30\%, or better, of {\it
planetary systems} composed of planets with well-sampled periods
($P\leq 5$ yr), and with the smaller component producing an
astrometric $S/N\geq 2$. Accurate coplanarity tests will be
possible for systems producing $S/N\geq 10$.

The frequency of multiple-planet systems, and their preferred
orbital spacing and geometry are not currently known. Based on
star counts in the vicinity of the Sun extrapolated from modern
models of stellar population synthesis, constrained to bright
magnitudes ($V< 13$ mag) and solar spectral types (earlier than
K5), we should expect $\sim 13\,000$ stars to 60 pc (Lattanzi et
al. 1999). GAIA, in its high--precision astrometric survey of the
solar neighborhood, will observe each of them, searching for
planetary systems composed of massive planets in a wide range of
possible orbits, making accurate measurements of their orbital
elements and masses, and establishing quasi-coplanarity (or
non-coplanarity) for detected systems with favorable
configurations. The size of the stellar sample is such to ensure
that the results obtained by GAIA would constitute a fundamental
complement to those which will come from other planet-search
programmes, in order to build the necessary statistics for deeper
theoretical understanding of planetary formation processes.


\end{document}